\documentclass[
    ,final            
  ]
  {aipproc}

\layoutstyle{6x9}

\newcommand{\ba}{\begin{array}{c}}
\newcommand{\ea}{\end{array}}
\newcommand{\nn}{\nonumber}

\newcommand{\be}{\begin{equation}}
\newcommand{\ee}{\end{equation}}
\newcommand{\chpt}{$\chi$PT}
\newcommand{\rcht}{R$\chi$T}
\newcommand{\ket}{\,\rangle}
\newcommand{\bra}{\langle \,}

\newcommand{\mF}{\mathcal{F}}

\newcommand{\mL}{\mathcal{L}}

\newcommand{\Frac}[2]{\frac{\displaystyle #1}{\displaystyle #2}}
\newcommand{\cO}{{\cal O}}

\begin{document}

\title{One loop predictions for the pion VFF\\
in Resonance Chiral Theory}



\author{J.J. Sanz-Cillero}{
  address={Grup de F\'\i sica Te\`orica and IFAE,
Universitat Aut\`onoma de Barcelona,\\
E-08193 Bellaterra (Barcelona), Spain}
}

\begin{abstract}
A calculation for  the one-loop pion vector form-factor
in Resonance Chiral Theory is provided in this talk.
The amplitude is computed  up to next-to-leading order in $1/N_C$ and,
by means of high-energy constraints,  we are able to produce a
prediction  for the corresponding   $\cO(p^4)$ Chiral
Perturbation Theory low energy constant
$L_9(\mu_0)=(7.6\pm 0.6)\cdot 10^{-3}$ at the
 scale  $\mu_0=770$~MeV.
\end{abstract}

\maketitle


\subsection{Introduction}

The issue of developing a quantum field theory for the
interaction of the  hadronic degrees of freedom
is still an open one.  More exactly, in this talk we focus
our attention on the description of
the  chiral Goldstones~\cite{chpt}
and the mesonic resonances.  We will work within a chiral invariant framework
for resonances, namely, Resonance Chiral Theory~\cite{RChTa,RChTb}.
The large--$N_C$ limit and the $1/N_C$ expansion will be taken
as guide lines to sort out the quantum field theory  computation,
implementing a perturbative counting
with the appropriate suppression of the hadronic
loops~\cite{NC}.

A pretty interesting observable to study is
the pion vector form-factor $\mF(q^2)$ (VFF):
\begin{eqnarray}
\bra \pi^+ \pi^- | \frac{1}{2}\bar{u}\gamma^\mu u
-\frac{1}{2}\bar{d}\gamma^\mu d |0\ket  &=&  (p_{\pi^+}-p_{\pi^-})^\mu
\, \mF(q^2)\, .
\end{eqnarray}
This amplitude is very well measured  experimentally,
being   a precise and basic  test for any proposed hadronic
description.
In order to provide reliable predictions for more complicated
observables~\cite{NLO-calc,RChTc,L10-Pich} one needs
to be able to describe well controlled QCD matrix elements as this, the
VFF~\cite{L9-Pich,prepara},
which  is known to be well dominated by the first
vector meson, the $\rho(770)$~\cite{Sakurai,VFF-Cillero}.

We will work within the single resonance approximation,
including only the chiral Goldstones and
the first resonance multiplets of vector  $(1^{--})$,
axial-vector $(1^{++})$,
scalar $(0^{++})$ and pseudo-scalar resonances
$(0^{-+})$~\cite{RChTa,RChTb,MHA}.
Likewise, only operators of at most
$\cO(p^2)$ are considered, i.e., with at most two derivatives~\cite{RChTa}.
Operators with a higher number of derivatives tend to violate
the high-energy behaviour prescribed by QCD~\cite{PI:08,NLO-satura}.
Likewise, the study of some particular amplitudes have shown
how these higher derivative terms of the Lagrangian
can be reduced into operators with a lower number
of derivatives and operators with only Goldstones  by means
of convenient meson field redefinitions~\cite{L9-Pich,S-pipi}.

Nothing restricts the number of meson fields in the operators of the
\rcht\ Lagrangian but for the organization of the one-loop computation
it is convenient to classify them  by their number of resonance fields:
\begin{eqnarray} \label{lagrangian}
\mathcal{L}_{R\chi T}&=&\mathcal{L}_G \,+\,\sum_{R}\mathcal{L}_{R}
\,+\,\sum_{R,R'}\mathcal{L}_{RR'}
\, + \, ... \,\,\,  ,
\nn
\end{eqnarray}
\vspace*{-0.75cm}
\begin{eqnarray}
\mL_G &=& \Frac{F^2}{4}\bra u^\mu u_\mu +\chi_+\ket \, ,
\nn\\
\mathcal{L}_R &=&
\frac{F_V}{2\sqrt{2}} \bra V_{\mu\nu} f^{\mu\nu}_+ \ket
+ \frac{i\, G_V}{2\sqrt{2}} \bra V_{\mu\nu} [u^\mu, u^\nu] \ket
+  \frac{F_A}{2\sqrt{2}} \bra A_{\mu\nu} f^{\mu\nu}_- \ket
+  i\,d_m \bra P \chi_- \ket
\nonumber \\
 &&\qquad +  c_d \bra S u_\mu u^\mu\ket + c_m\bra S\chi_+\ket \,,
\label{eq.LG+LR}
\end{eqnarray}
provided in Ref.~\cite{RChTa}, together with
their corresponding kinetic terms $\mL[R]^{Kin}$.
In addition, one has several other
operators with two resonance fields that may
enter in the VFF  at next-to-leading order in $1/N_C$
(NLO)~\cite{L10-Pich,L9-Pich}.
Nonetheless,  only the diagrams with a cut
of two Goldstones or a Goldstone and a resonance
will be taken into account, being the absorptive channels with
two   resonances kinematically
suppressed~\cite{L10-Pich}. The relevant vertices
that  may then contribute to the VFF at one
loop are~\cite{RChTc,L10-Pich,prepara}:
\begin{eqnarray}
 \mathcal{L}_{SA} &=&
   \lambda^{SA}_1 \bra \{\nabla_\mu S, A^{\mu\nu} \} u_\nu \ket
 \, ,
%
\quad
 \mathcal{L}_{PV} =
   i \lambda^{PV}_1\bra [\nabla^\mu P,V_{\mu\nu} ] u^\nu \ket
 \, ,
%
\\
 \mathcal{L}_{SP} &=&
 \lambda^{SP}_1\bra u_\alpha \{\nabla^\alpha S,P\} \ket \, ,
%
\qquad
 \mathcal{L}_{VA}\!\!=\!\!  i \lambda^{VA}_2 \! \bra \!  [ V^{\mu\nu}, A_{\nu\alpha} ] h^\alpha_\mu \!\ket  \! +\! i \lambda^{VA}_3 \! \bra \! [ \nabla^\mu V_{\mu\nu}, A^{\nu\alpha} ] u_\alpha \!\ket   \nonumber  \\
&&\hspace*{4.cm}+ i \lambda^{VA}_4 \bra
[ \nabla_\alpha V_{\mu\nu}, A^{\alpha\nu} ] u^\mu \ket
+ i \lambda^{VA}_5 \bra  [ \nabla_\alpha V_{\mu\nu}, A^{\mu\nu} ]
u^\alpha \ket
\,.
\nn
\end{eqnarray}
The brackets $\langle ... \rangle$ denote   trace in
flavour space and the  chiral tensors $u^\mu$, $\chi_\pm$,
$f_\pm^{\mu\nu}$  (containing Goldstones and external sources)
are defined  in Refs.~\cite{RChTa,RChTc}.
%
In this talk we refer to the diagrammatical
quantum field theory calculation of the VFF at NLO although
its derivation  through dispersion relations
is completely equivalent and can be found in Ref.~\cite{prepara}.
Further details on the $\cO(p^6)$ LEC predictions and
alternative numerical estimates  can be found there.
The chiral limit is assumed all along the talk.

\vspace*{-0.75cm}
\subsection{High-energy conditions}

The full VFF is well known to vanish when
$q^2\to\infty$~\cite{Brodsky-Lepage}.   Thus,
\rcht\ can be  then used as an interpolator between both regimes,
showing  at leading order (LO)   the simple structure~\cite{RChTb,MHA}
\begin{equation}
\mF(q^2) \, =\, 1\, +\, \Frac{F_V G_V}{F^2}
\Frac{q^2}{M_V^2 -q^2}
\quad \stackrel{F_VG_V=F^2}{=}\quad \Frac{M_V^2}{M_V^2-q^2}\, ,
\label{eq.VFF-LO1}
\end{equation}
where the requirement
that the VFF vanishes at $q^2\to\infty$ leads to the
LO relation ${  F_V G_V=F^2   }$~\cite{RChTb}
and the usual monopolar form for the VFF.
This expression  can be also understood from a Pad\'e--approximant
point of view as a $[0/1]$ Pad\'e--type approximant with
the pole fixed to $M_\rho^2$~\cite{VFF-Pade},
being the first of a series  of Pad\'e sequences.

At NLO in $1/N_C$,
the corresponding one-loop diagrams are
ultraviolet (UV) divergent~\cite{NLO-calc,L9-Pich,Rosell-genera,VFF-RGE}  and,
in addition to the renormalization of some couplings
of the LO Lagrangian,  one needs to introduce
some subleading operators~\cite{L9-Pich,VFF-RGE},
\begin{eqnarray}
\mathcal{L}^{\rm G}_{\rm NLO} &=&\,
-\, i\, \widetilde{L}_9\bra f_+^{\mu\nu} u_\mu u_\nu\ket \, ,
\label{eq.L9t}
\\
\mathcal{L}^V_{\rm NLO} &=&\,
X_Z\bra V_{\lambda\nu}\nabla^\lambda\nabla_\rho \nabla^2 V^{\rho\nu}\ket
\,+\,  X_F\bra V_{\mu\nu}\nabla^2 f_+^{\mu\nu}\ket
+ \,2\, i\, X_G    \bra V_{\mu\nu} \nabla^2 [u^\mu, u^\nu] \ket
\,.
\nn
\end{eqnarray}
However,  the $\mathcal{L}^V_{\rm NLO}$
couplings $X_{Z,F,G}$ are not physical by themselves:
it is not possible to fix them univocally from the experimental VFF.
Indeed, as these subleading $\mL^V_{\rm NLO}$ operators
are proportional to the equations of motion, one finds that
$\mathcal{L}^V_{\rm NLO}$ can  be fully transformed
into the $M_V$, $F_V$, $G_V$ and $\widetilde{L}_9$ terms  and
into other operators that  do not contribute
to the VFF by means of meson field redefinitions~\cite{L9-Pich,VFF-RGE}.
Thus, the on-shell VFF does not really need  all the terms
in Eq.~(\ref{eq.L9t})
to make the amplitude finite, just $\widetilde{L}_9^{\rm eff}$,
$F_V^{\rm eff} G_V^{\rm eff}$ and
$M_V^{{\rm eff}\,\, 2}$~\cite{L9-Pich,VFF-RGE}.
In what follows, we will always refer to the simplified Lagrangian and
the ``eff'' superscript of the LO parameters  will be implicitly assumed.

As we did before at large--$N_C$, we can now take the
one-loop VFF and use it as an interpolator between high and low energies,
by imposing again short-distance constraints on $\mF(q^2)$.
In a similar way, its spectral function
Im$\mF(q^2)$ must go to zero at high energies.
In the present work~\cite{prepara},  we will actually
impose this constraint channel by channel, i.e., we will demand
that each separate two-meson cut    Im$\mF(q^2)|_{M_1,M_2}$   vanishes
at $q^2\to\infty$.  Actually, for spin--0 mesons this must be so as
its one-loop contribution to the spectral function is essentially
the VFF at LO (which vanishes at infinite momentum)
times the partial-wave scattering amplitude at LO (which
is upper bounded).  For  the higher spin resonances
the derivation is more cumbersome as the Lorentz structure
allows for the proliferation of form-factors and
the unitarity relations are not that simple. Still, in many situations
it has been already found that these  amplitudes with
massive spin--1 mesons as final states must  go to zero
at high energies
even faster due to the presence of extra powers
of momenta in the unitarity relations coming from intermediate
longitudinal polarizations~\cite{L10-Pich}.
%

The high-energy expansion of our one-loop
\rcht\ expression yields  the   structure
\begin{eqnarray}
\mbox{Im}\mF(q^2) &=&
 q^2\,\left(\beta_2^{(p)}
+\beta_2^{(\ell)}\ln\Frac{-q^2}{M^2} \right)
  +
\left(\beta_0^{(p)}
+\beta_0^{(\ell)}\ln\Frac{-q^2}{M^2} \right)
 +  \cO\left(\Frac{1}{q^2}\right)\, ,
\end{eqnarray}
which requires the constraints
$\beta_2^{(p)}=\beta_2^{(\ell)}=\beta_0^{(p)}=\beta_0^{(\ell)}=0$.
The $\ln(-q^2/M^2)$ terms are produced by the triangle diagrams
with crossed exchanges of resonances of mass $M$.
%
%
The short-distance conditions derived from every channel are:
\begin{itemize}
\item
\underline{$\pi\pi$ channel}:

\vspace*{-0.95cm}
\begin{equation}
F_V G_V\,=\, F^2 \,, \qquad     3\,G_V^2+2\,c_d^2 \,=\, F^2
 \,, \label{constraintpipi}
\end{equation}

where the first one coincides with the large--$N_C$ constraint
for the VFF. The second one is consistent
with that obtained in the context of the $\pi\pi$--scattering
at LO~\cite{Guo-aIJ}.

\item
\underline{$P\pi$ channel}:

\vspace*{-1.25cm}
\begin{eqnarray}
\lambda_1^{\mathrm{PV}}\,=\,0\,, \label{constraintPpi}
\end{eqnarray}
consistent with the large--$N_C$ constraint from the
vector form-factor into $P\pi$,
studied  in Ref.~\cite{L10-Pich}.
This kills completely the $P\pi$ loop contribution
to the $\pi\pi$ VFF.

\item
\underline{$A\pi$ channel}:
The constraints have several solutions   but
we have kept just those consistent with
the large--$N_C$ vector form-factor into $A\pi$,
studied in Ref.~\cite{L10-Pich}.
%

\vspace*{-0.5cm}
\begin{eqnarray}
-2 \lambda_2^{\mathrm{VA}} + \lambda_3^{\mathrm{VA}}&=& 0 \,,
\phantom{\frac{1}{2}} \nonumber \\
- \lambda_3^{\mathrm{VA}} + \lambda_4^{\mathrm{VA}}
+ 2 \lambda_5^{\mathrm{VA}} &=& \frac{F_A}{F_V} \,, \nonumber \\
-\frac{F_A \,G_V\left(M_A^2 - 4\,M_V^2\right)}{
3\sqrt{2} M_A^2 c_d\,F_V}&=& \lambda_1^{\mathrm{SA}}
\,. \label{constraintApi}
\end{eqnarray}

\end{itemize}

After imposing the right high-energy behaviour on the spectral function
the logarithmic  and polylog terms of the VFF also
vanish at $q^2\to\infty$ and only the purely rational part
has the wrong behaviour.
The one-loop contribution  has a unique decomposition in the
form~\cite{L10-Pich,prepara}
\begin{eqnarray}
\mF(q^2)^{1-\ell oop} &=&
\overline{\mF}(q^2)^{1-\ell oop} \, +\, \Frac{2 q^2}{F^2} \hat{\delta}_2
\,+\, \hat{\delta}_0 \Frac{q^2}{M_V^2-q^2}
\,+\, \hat{\delta}_{-2}\Frac{q^2}{(M_V^2-q^2)^2}\, ,
\label{eq.1loop+disp}
\end{eqnarray}
where the subtracted function $\overline{\mF}(q^2)^{1-\ell oop}$
can be obtained through a once-subtracted dispersion relation and it
is fully determined by the two--meson
spectral function Im$\mF(q^2)$~\cite{L10-Pich,prepara}.
It behaves like $\cO(q^0)$ at high energies and has no contribution
to the real part of the single and double poles at $q^2=M_V^2$,
which are fully given by $\hat{\delta}_0$ and $\hat{\delta}_{-2}$.
The UV divergences are contained in the real constants
$\hat{\delta}_k$.  Actually, we will consider the
on-shell vector mass scheme $\delta M_V^2$ such that the real part
$\hat{\delta}_{-2}$ of  the double pole is completely removed.
The form-factor has then the structure~\cite{L9-Pich,prepara}
\begin{eqnarray}
\mF(q^2) &=& 1\,+\, \left(\Frac{F_V G_V}{F^2}+\hat{\delta}_0\right)
\Frac{q^2}{M_V^2-q^2}
\,+\, \Frac{2 q^2}{F^2}\left(\widetilde{L}_9 +\hat{\delta}_2\right)
\, +\,  \overline{\mF}(q^2)^{1-\ell oop}\, ,
\end{eqnarray}
where the subtracted loop contribution behaves at high energies like
$   \overline{\mF}(q^2)^{1-\ell oop}=\delta_0 +\cO(1/q^2)     $
and leads to the VFF expansion,
\begin{eqnarray}
\mF(q^2) &=&
\Frac{2 q^2}{F^2}\left(\widetilde{L}_9 +\hat{\delta}_2\right)
\,+\,
\left( 1+\delta_0\,-\, \Frac{F_V G_V}{F^2}- \hat{\delta}_0
 \right)
\,+\,  \cO\left(\Frac{1}{q^2}\right) \, .
\end{eqnarray}
After demanding now that the VFF  vanishes as $q^2\to\infty$,
one gets the NLO constraints
\begin{eqnarray}
\widetilde{L}_9 +\hat{\delta}_2\, =\, 0\, ,
\qquad \qquad
\Frac{F_V G_V}{F^2}+  \hat{\delta}_0
\,=\, 1+\delta_0\, .
\label{eq.NLO-L9}
\end{eqnarray}
The subleading corrections $\hat{\delta}_2$ and $\hat{\delta}_0$
will always appear in combination with
$\widetilde{L}_9$ and $F_V G_V/F^2$, respectively,
which absorb their UV divergence.
The expressions from Eq.~(\ref{eq.NLO-L9})
can be compared to their
large--$N_C$ values $\widetilde{L}_9=0$ and $F_V G_V/F^2=1$.
Thus, the  $\mu$--independent constant
$\delta_0$ is the actual relevant quantity here, which will
ultimately participate in the LEC determination.
In Table~\ref{tab.delta0} one can find the contributions from
the various channels. We also provide its final
contribution to the chiral LEC,  $L_9(\mu)=... +  \frac{F^2}{2 M_V^2}
\delta_0$,   as we will see in the next section.
After considering the relations (\ref{constraintpipi}),
(\ref{constraintPpi}), (\ref{constraintApi}) and (\ref{eq.NLO-L9})
the spectral functions
can be expressed in terms of $G_V$, $F_A$, $F$ and masses.

\begin{table}
\begin{tabular}{cccc}
\hline
  &\tablehead{1}{r}{b}{$   \pi\pi \quad $  }
  & \tablehead{1}{r}{b}{$  A\pi \quad $  }
  & \tablehead{1}{r}{b}{$  P\pi   $ }
  \\
\hline
$\delta_0\qquad \,\,\,\,$ &   $ \quad  \quad 0.23$  &   $  \quad \quad 0.14$
&    $ \quad  \quad 0$
  \\
$\frac{F^2}{2 M_V^2} \delta_0\qquad$ &  $  \quad \quad 1.5\cdot 10^{-3}$   &
$ \quad  \quad 1.0\cdot 10^{-3}$  &       $  \quad \quad 0$
\\
\hline
$\xi_{L_9}\qquad$ &  $  \quad \quad -1.6\cdot 10^{-3}$   &
$ \quad  \quad -0.1\cdot 10^{-3}$  &       $  \quad \quad 0$
\\
\hline
\end{tabular}
\caption{{\small
Correction $\delta_0$ to the renormalized combination of couplings
$F_V G_V/F^2$ from the different two--meson channels.
The low-energy contribution $\xi_{L_9}$ from the one-loop
part $\overline{\mF}(q^2)^{1-\ell oop}$ is also provided.
}}
\label{tab.delta0}
\end{table}

\vspace*{-0.75cm}
\subsection{Low-energy expansion and predictions}

The low-energy expansion of the one-loop part produces the massless
\chpt\  log together with a series of analytical terms:
$\overline{\mF}(q^2)^{1-\ell oop}=\frac{2 q^2}{F^2} \xi_{L_9}
+\frac{q^2}{F^2}\frac{\Gamma_9}{16\pi^2}\left(
\frac{5}{3}-\ln\frac{-q^2}{\mu^2}
\right)+\cO(q^4)$, where part of the $\pi\pi$ loop contribution has been
explicitly separated of $\xi_{L_9}$ for convenience for the matching
with \chpt .   The \rcht\ coefficient that appears
in front of the chiral log
is exactly $\Gamma_9=1/4$~\cite{chpt}, ensuring the recovery of the
proper renormalization scale dependence of the LEC.
Thus, independently of the value of the \rcht\ parameters
the chiral symmetry invariance allows one always
to match the low-energy \chpt\ expression~\cite{VFF-chpt,L9-Bijnens},
\vspace*{-0.5cm}
\begin{eqnarray}
\mF(q^2) &=& 1 +\Frac{ 2 L_9(\mu) q^2}{F^2}
+\Frac{\Gamma_9}{32\pi^2}
\left(\Frac{5}{3}-\ln\Frac{-q^2}{\mu^2}\right)  +\cO(q^4) \, .
\end{eqnarray}
Substituting  the short-distance constraints from the previous section,
one gets  the simple form for the LEC prediction,
\vspace*{-0.5cm}
\begin{eqnarray}
L_9(\mu)\,=\, \Frac{F^2}{2 M_V^{ 2}}
\,\Big(  \,1\,+\,\delta_0\,\Big)
\,\,+\,\,  \xi_{L_9}(\mu)\, .
\end{eqnarray}
For illustrative reasons we provide in Table~\ref{tab.delta0}
the numerical contributions from the different two--meson channels
to $\xi_{L_9}$.

At large $N_C$, one has the LO estimate
$L_9 =\frac{F_V G_V}{2 M_V^2 }=\frac{F^2}{2 M_V^2}\simeq
6.8\cdot 10^{-3}$~\cite{RChTb}.  This determination however lacks of
the one-loop \chpt\ running, so it carries an uncertainty on the saturation
scale which can be naively estimated as
$\Delta L_9\simeq 0.5\cdot 10^{-3}$ by
varying $\mu^2$ between $M_\rho^2/2$
and $2 M_\rho^2$~\cite{RChTa,Guo-aIJ}.
We will compare this to the LEC prediction at NLO
in $1/N_C$ with the inputs $M_V=770\pm 5$~MeV,
$M_S=1090\pm 110$~MeV, $F=89\pm 2$~MeV,
with $G_V$ varied between its limit upper
value $F/\sqrt{3}$ and $40$~MeV,
$M_A=1200\pm 200$~MeV and $F_A=120\pm 20$~MeV.
If we add the one-loop diagrams with $\pi\pi$ absorptive cut,
one obtains $L_9(\mu_0)=(6.6\pm 0.4) \cdot 10^{-3}$
for the standard comparison scale $\mu_0=770$~MeV.
Finally, if the $A\pi$ channel is also added (the $P\pi$ one is exactly
zero after the high-energy constraints),
we get the final prediction provided in~Table~\ref{tab.L9-bib},
where it is compared to previous
determinations~\cite{chpt,VFF-Cillero,L9-Bijnens,L9-Prades}.


\begin{table}
\begin{tabular}{lccccccc}
\hline
  & \tablehead{1}{c}{b}{$\cO(p^4)$ \chpt\ \\
  \cite{chpt} }
  & \tablehead{1}{c}{b}{$\cO(p^6)$ \chpt\ \\
   \cite{L9-Bijnens}   
 }
  & \tablehead{1}{c}{b}{ $\cO(p^4)$ $\tau$--SR
  \\ \cite{L9-Prades}}
  & \tablehead{1}{c}{b}{ $\cO(p^6)$ $\tau$--SR
  \\ \cite{L9-Prades}}
  & \tablehead{1}{c}{b}{ $\tau$--\rcht\  \\
  \cite{VFF-Cillero}  }
  & \tablehead{1}{c}{b}{ This work \\
     \cite{prepara}  }
  \\
\hline
$10^3\, L_9(\mu_0)$ & $6.9\pm 0.7$  & $5.93\pm 0.43$
& $6.54\pm 0.15$  & $5.50\pm 0.40$
 &   $7.04\pm 0.23$  &    $7.6\pm 0.6$
\\
\hline
\end{tabular}
\caption{{\small
Comparison of  our $L_9(\mu_0)$ determination at $\mu_0=770$~MeV with previous analyses.
}}
\label{tab.L9-bib}
\end{table}


\vspace*{-0.75cm}
\subsection{Acknowledgements}

\vspace*{-0.35cm}
{\small
Talk given at
QCD@Work 2010 --International Workshop on QCD: Theory and Experiment--,
20-23 June 2010,  	Martina Franca, Valle d'Itria  (Italy).
I would like to thank the organizers for their attentions
during the conference.
This work
was supported in part by
the Ministerio de Ciencia e Innovac\'\i on
under grant CICYTFEDER-FPA2008-01430,
the Juan de la Cierva Program,
the EU Contract No. MRTN-CT-2006-035482 --``FLAVIAnet''--,
the Spanish Consolider-Ingenio 2010 Programme CPAN (CSD2007-00042)
and the Generalitat de Catalunya under grant SGR2009-00894.
}




\bibliographystyle{aipproc}   




\end{document}